\documentclass[prab,10pt,twocolumn,showpacs,superscriptaddress,showkeys]{revtex4-2}

\usepackage{graphicx}
\usepackage{dcolumn}
\usepackage{bm}
\usepackage{amssymb}
\usepackage{amsmath}
\usepackage{array}
\usepackage{rotating}
\usepackage{color}
\usepackage{tabularx}
\usepackage{hyperref}
\usepackage[UKenglish]{babel}
\usepackage[utf8x]{inputenc}
\usepackage[T1]{fontenc}
\usepackage[dvipsnames]{xcolor}

\begin{document}

\title{Nanometre-scale emittance beams from a continuous-wave RF gun}

\author{Anatoliy Opanasenko}
\affiliation{NSC/KIPT, Akademicheskaya 1, Kharkiv 61108, Ukraine}

\author{Zolt\'an Tibai}
\affiliation{Institute of Physics, University of P\'ecs, P\'ecs 7624, Hungary}

\author{K\'evin Pepitone}
\affiliation{Uppsala University, L\"agerhyddsv\"agen 1, Uppsala 75120, Sweden}

\author{Georgii Shamuilov}
\email[e-mail: ]{georgii.shamuilov@physics.uu.se}
\affiliation{Uppsala University, L\"agerhyddsv\"agen 1, Uppsala 75120, Sweden}

\author{Vitaliy Goryashko}
\email[e-mail: ]{vitaliy.goryashko@physics.uu.se}
\affiliation{Uppsala University, L\"agerhyddsv\"agen 1, Uppsala 75120, Sweden}

\begin{abstract}
    The operation of Ultrafast Electron Diffractometers (UEDs) and Free-Electron Lasers (FELs) relies on high-brightness electron beams produced by radio-frequency (RF) photocathode guns. The next generation of high-repetition rate UEDs and FELs requires electron beams with a high average brightness. To this end, we introduce a continuous wave RF photocathode gun at 325 MHz with an APEX-like geometry. The gun allows for the production of electron beams with very high both peak and average 5D~brightness while having moderate RF power consumption. The gun is operated in blowout regime with an energy gain of 0.4~MeV and a peak cathode field of 35~MV/m. Via massive numerical simulations, we exemplify three regimes of the gun operation: (i) 160~fC electron beams with a 5-nm-scale emittance for UEDs, (ii) 1.6~pC beams with a 20-nm-scale emittance for table-top FELs and dielectric-based accelerators, and (iii) 16~pC beams with a 50-nm-scale emittance for inverse Compton sources and other accelerator-based photon sources. We introduce a simple analytical model for the formation of the virtual cathode~-- the onset of the suppression of photoemission current due to space-charge forces. The model accounts for the laser pulse duration. Furthermore, our extensive numerical simulations indicate a well-pronounced maximum in the 5D beam brightness for the laser spot radius approximately 150\% of that corresponding to the onset of the virtual cathode. The finding does not support the common approach in the literature that in the blowout regime the laser spot radius must be much larger than the critical radius corresponding to the virtual cathode onset. 
    \end{abstract}


\date{\today}
\maketitle


\section{Introduction}

A sustainable transition to green economy requires new solutions to harness, store and transport energy~\cite{EU_commission}. In its turn, this requires deeper understanding of energy transfer in materials and chemical reactions. Ultrafast Electron Diffractometers (UEDs) and Free-Electron Lasers (FELs) are cutting-edge tools for studying the dynamics of nonequilibrium states of and revealing energy flows in matter with picometre spatial and femtosecond temporal resolution (note: FELs can provide even sub-fs resolution) \cite{Sciaini2011,Miller2014,Emma2010_all_authors,Mak2019,Salen2019,Ishikawa2012,Huang2017,Duris2019}. The operation of UEDs and FELs relies on high-brightness electron beams serving either directly as a probe for studying matter in UED instruments or as a source of photon flashes of X-ray radiation in FELs. The quality of electron beams -- the 6D brightness -- defined as the number of electrons per unit 6D phase space sets the spatial, angular and temporal resolution of UEDs and the brightness of FELs.


Normal conducting, Mega-electron-Volt (MeV) RF electron guns produce electron beams of the highest brightness and are at the heart of X-ray FELs and of the next generation UEDs \cite{Bazarov2009,Li2012,Weathersby2015_all_authors,Filippetto2016,Simone_rf_gun}. However, MeV RF guns are typically pulsed at 100 Hz and while the peak brightness of electron beams is enormous, the average brightness is almost negligible. This limitation of low-repetition-rate operation arises from the resistive RF heating of the gun cavities, while cooling systems are limited in size and interface area. Yet, it is desirable to have MHz-scale beam repetition rates because the time needed to gather data for a given experiment decreases correspondingly by orders of magnitude. In addition, the facility wall-plug power consumption for a given experiment greatly reduces. 

To operate at higher repetition rates of the order of one MHz or higher and increase the average brightness of produced electron beams, several types of CW photocathode guns have been proposed: (i) DC guns, (ii) superconducting RF guns, (iii) very-high-frequency (VHF) RF guns and (iv) hybrid guns using a combination of a DC gun followed by a SC cavity~\cite{Teichert_2011,qian2019overview}. From all these electron guns, the VHF RF photo-gun is particularly advantageous for employment in user facilities due to its high performance and high reliability. In particular, the VHF RF photo-gun APEX developed at Lawrence Berkeley National Laboratory~\cite{Sannibale2012,Sannibale2019} is employed in the next generation FEL, LCLS-II~\cite{APEX_LCLS}. 

In this paper, we introduce a CW RF photocathode gun in the ultrahigh frequency (UHF) range (325 MHz). The gun is operated with an accelerating field of 35~MV/m  on the cathode. It provides an energy gain of 0.4~MeV for a moderate RF power consumption of around 20~kW CW, Sec~II. The  RF power consumption is reduced by a factor of four compared to the APEX gun but without compromising on beam quality. The optimised geometry of the gun cavity, Sec~II, provides a very uniform accelerating field with a negligible RF-induced emittance growth. For different parameters of the initial bunch distribution,  massive numerical simulations are carried out, Sec.~III. The regimes of blowout generation with a nearly thermal-level emittance are demonstrated in a wide range of bunch charge values from 160~fC to 16~pC. The simulations predict an unexpected result that the lowest beam emittance is achieved when the space-charge field on the cathode, $E_\mathrm{sp-ch}$, is almost 50\% of the acceleration field, $E_\mathrm{acc}$. This is in a contrast with the common approach in the literature that the condition $E_\mathrm{sp-ch} \ll E_\mathrm{acc}$ must be met in blowout regime~\cite{Luiten2004,Musumeci2008,vanOudheusden2010}. We also derive a simple analytical estimate for the onset of virtual cathode formation depending on the laser spot size and duration, Sec.~IV. A comparison to the state-of-the-art photocathode guns, Sec.~V, clearly shows that the proposed gun outperforms the existing guns in terms of emittance for bunch charges of a few tens of pC. Hence, the proposed gun might be a good solution for applications requiring high-brightness low-charge electron beams.

\section{RF design and performance of the gun cavity}

The gun cavity is designed to operate at 325~MHz with a threefold purpose: 
\textbf{(i)}~to take advantage of the recent development of high-efficiency, high-power solid-state amplifiers in the MHz frequency range~\cite{marchand2007high,di2009solid,jacob2015radio,Dragos_2014}, 
\textbf{(ii)}~to have the gun frequency compatible with that of the TESLA cavity (1.3 GHz) for further beam acceleration and manipulation, and \textbf{(iii)}~to make the cavity compact compared to lower frequency counterparts. 

\begin{figure}[tb]
\centering
\includegraphics[width = \linewidth]{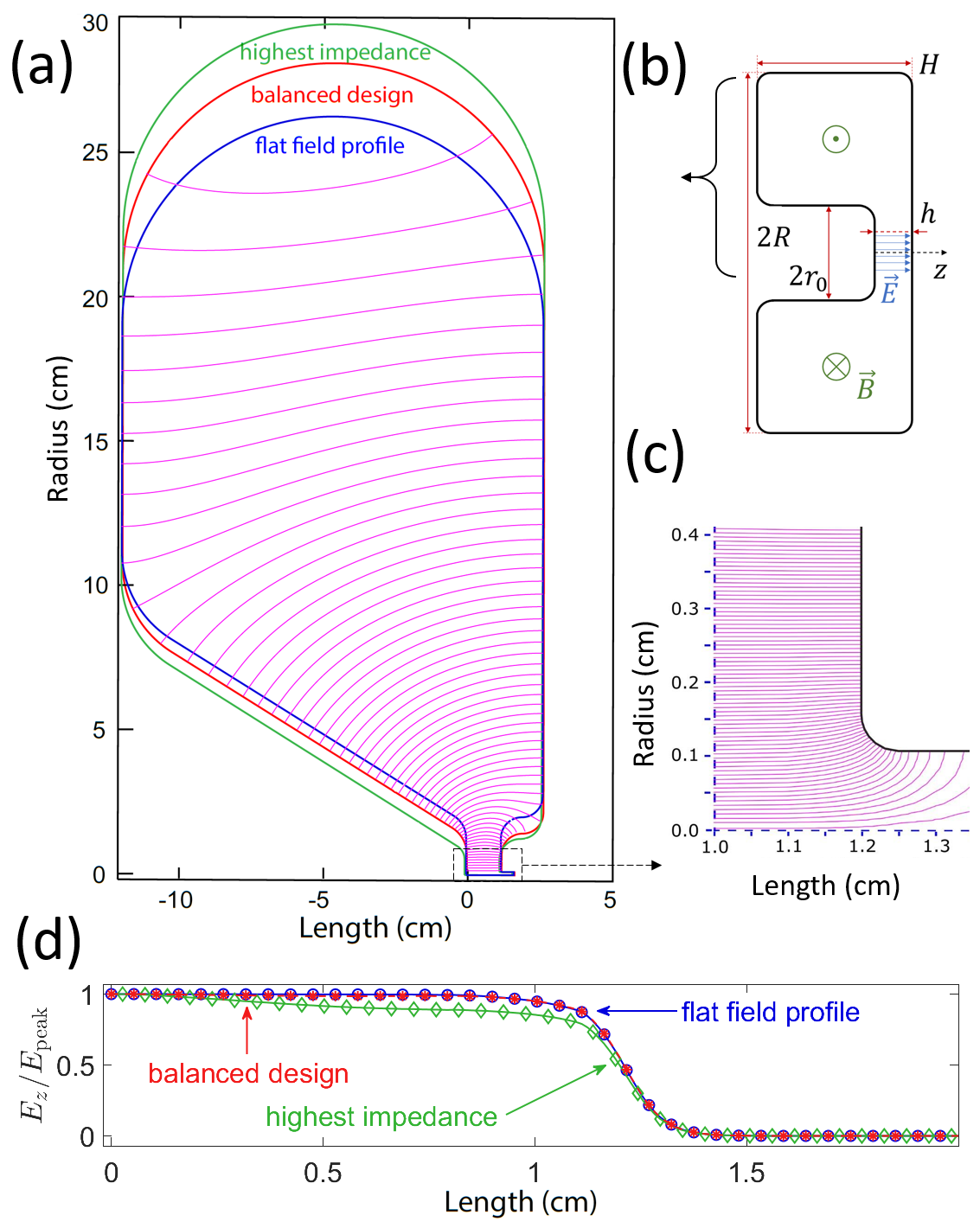}
\caption{\textbf{(a)} Geometries of the three cavity variants (rotational symmetry). \textbf{(b)} Representation of the simplest re-entrant cavity geometry with relevant dimensions. \textbf{(c)} Zoom in into the region of the output aperture. \textbf{(d)} Electric field distributions along the gun axis for the three cavity variants.}
\label{fig:Gun_geometry}
\end{figure}

As part of the strategy of making the gun cavity compact, the re-entrant cavity geometry (widely adopted in klystrons~\cite{fujisawa1958general}) is used, see Fig.~\ref{fig:Gun_geometry}a. Recall that for simple cavity geometries, the wavelength of the lowest eigenmode, $\lambda_{res}$, is roughly equal to the largest dimension of the cavity, $D$. The re-entrant cavity has no such limitation and $D \ll \lambda_{res}$ is possible. 
The re-entrant cavity is formed by a capacitor-like accelerating region and by an inductive toroidal part, see  Fig.~\ref{fig:Gun_geometry}b. The electric field of the lowest mode is mostly concentrated in the accelerating region whereas the magnetic field is mostly in the toroidal part. 

\begin{table*}[tb]
\centering
\begin{tabular}{lccc}
\hline \hline
Parameters & Low power & Flat field & Balanced \\
\hline
Quality factor $Q_0$ & 44890  & 36600   & 41210 \\
Shunt impedance, M$\Omega$ & 8.7  & 6.6   & 7.8 \\
Shunt impedance per unit length, M$\Omega$/m & 529.1  & 400.7   & 471.3 \\
Peak-to-cathode field ratio $E_\mathrm{max}/E_\mathrm{cat}$ & 1.3  & 1.2   & 1.2 \\
Electric field at the cathode, MV/m & 35  & 35   & 35 \\
Energy gain, keV & 389  & 419   & 417 \\
Power dissipation, kW & 17.3  & 26.6   & 22.4 \\
Max. power density on the wall, W/cm$^2$ & 13.7  & 23.5   & 18.6 \\
Stored energy, J & 0.42  & 0.48   & 0.45 \\
\hline \hline
\end{tabular}
\caption{RF parameters for the three chosen cavity designs.}
\label{tab:RF_parameters}
\end{table*}

The shape of the re-entrant cavity is optimised with respect to the following three objectives: \textbf{(i)}~maximal shunt impedance [minimal RF power dissipation, $R_\mathrm{sh}~=~V^2/2P_\mathrm{loss}$, where $V$ is the accelerating voltage and $P_\mathrm{loss}$ is the power dissipation]; \textbf{(ii)}~maximal electric field on the cathode [rapid acceleration mitigates the emittance growth due to space-charge]; \textbf{(iii)}~maximal flatness of the accelerating field near the axis [minimal nonlinear terms in the transverse field $E_r(r)$ to reduce an RF-induced correlated emittance growth].

To facilitate the analysis of the RF performance and the optimisation of the cavity, it is convenient to start with the re-entrant cavity in its simplest form, shown in Fig.~\ref{fig:Gun_geometry}b. 
In a quasistatic approximation (the cavity dimensions are smaller than the wavelength in question),
the capacitance and inductance of the cavity are estimated as
$C = r_0^2/4h$ and $L = (2H/c^2) \ln(R/r_0)$, respectively.  Here, $R$ and $r_0$ 
are the radii of the cavity and post, respectively; 
$H$ is the height of the cavity and $h$ is the length of the acceleration gap.
Then, the resonant wavelength reads:
\begin{equation}\label{lambda_r}
 \lambda_r \equiv 2\pi c \sqrt{LC} = \frac{2\pi R}{\sqrt{2h/H}} \, \frac{\sqrt{\ln R/r_0}}{R/r_0}.
\end{equation}

To have a uniform accelerating field in the gap (flat field profile), 
the radii of the noses in Fig.~\ref{fig:Gun_geometry}a must be much larger than 
the accelerating gap length. One can, however, realise that this contradicts the requirement of a maximal impedance. Let us discuss this in some details using the above analytical model. 
First, we note that the energy of the electric field, $W_E = CV^2/2$, 
increases with $r_0$ for a fixed accelerating voltage $V$. Second, for an eigenmode the energy of the magnetic field, $W_H = LI^2/2$, is equal to that of the electric field. Hence, a larger $W_E$ implies a larger current $I$ in the walls for a fixed impedance $L$ (fixed geometry of the toroidal part). Therefore, a larger $r_0$ implies higher RF losses, \textit{i.e.} a lower impedance. 

On the other hand, the inner radius $r_0$ can be increased without raising losses by increasing the inductance $L$ via increasing the cavity height $H$. But even so, a room for increasing $L$ has a limit defined by the requirement of compactness. In addition, the condition for the resonant wavelength  $\lambda_r = 2\pi c \sqrt{LC}$ must be met. Clearly, the objectives are conflicting and any design is a balance between the requirements.  

We carry out the cavity design using SUPERFISH code~\cite{Superfish} and present here three versions of the cavity geometry to illustrate how the main geometrical parameters affect the RF cavity performance. The version (1) has the highest shunt impedance, the version (2) is optimised for a flat field profile, and the version (3) is a balanced design between the two. The  cavity shapes and electric field distributions are shown in Fig.~\ref{fig:Gun_geometry} while their main parameters are summarised in Table~\ref{tab:RF_parameters}.

As we already indicated for the simplest re-entrant cavity, the electric field is mainly concentrated in the capacitor-like accelerating gap between the conical noses, Fig.~\ref{fig:Gun_geometry}a,~c. Increasing the radii of the noses provides a more flat electric field profile, see Fig.~\ref{fig:Gun_geometry}d, but at the expense of a lower shunt impedance of the cavity. For gun operation, we chose the balanced version of the geometry (red curve in Fig.~\ref{fig:Gun_geometry}), as a compromise between the two extremes of the maximal shunt impedance and of the most flat field profile.

\section{Electron beam dynamics}

Apart from the field profile given by the shape of the cavity, the initial beam distribution is of utmost importance. Directly after emission, having a sub-relativistic velocity, electrons in the bunch repel (space-charge), which makes emittance and brightness deteriorate. Properly choosing transverse and longitudinal distributions of the bunch allows us to partly mitigate this emittance growth.
Unlike other emission mechanisms, photoemission provides a large degree of control over electron bunches. Properties of the driving laser pulse (waist size, duration, profile, pulse energy etc.) allow for a flexible control  of the electron bunch properties (transverse size, duration, distributions, bunch charge, respectively). Though, differences also arise between the laser pulse and electron bunch   because electrons suffer from space-charge fields. Thus, the brightness of electron bunches greatly depends on the chosen parameters of the laser pulse.

Photoemission can be utilised in two regimes: (i)~steady-state emission with the maximum current density given by Child-Langmuir law~\cite{Child1911,Langmuir1923,Luginsland1996,Lau2001,Filippetto2014,Shamuilov2018}, and (ii)~fundamentally transient blowout regime for the creation of uniformly filled ellipsoidal bunches~\cite{Luiten2004,Musumeci2008,vanOudheusden2010}.
For the former, the bunch aspect ratio is usually large, $L/R\gg 1$ (long cylindrical bunches), while for the latter it is the opposite, $L/R\ll 1$ (pancake-like bunches). Long bunches retain their shape as they propagate through the RF gun and further. On the contrary, pancake-like bunches evolve into uniformly filled ellipsoids ("waterbag") with linear space-charge forces. We employ blowout regime since it imposes no requirements on the longitudinal laser pulse profile -- the laser duration must be just short, typically 30~fs.  The transition between $L/R\ll 1$ and $L/R\gg 1$ is not instant. As we shall see in the simulations, an intermediate regime exists, but leads to degradation of the beam quality.

In our studies we use established numerical codes for space-charge-dominated beam dynamics simulations, namely ASTRA~\cite{astra_code}, RF-track~\cite{RFtrack} and GPT~\cite{GPT}. The results are cross-checked between the three codes and the presented figures are produced using ASTRA. In the simulations, we assume a parabolic form of the transverse distribution of the laser pulse intensity on the cathode and the initial electron bunch distribution is generated from the Fermi-Dirac statistics using the procedure described in~\cite{Dowell2009}.

\begin{figure*}[tb]
\centering
\includegraphics[width = \linewidth]{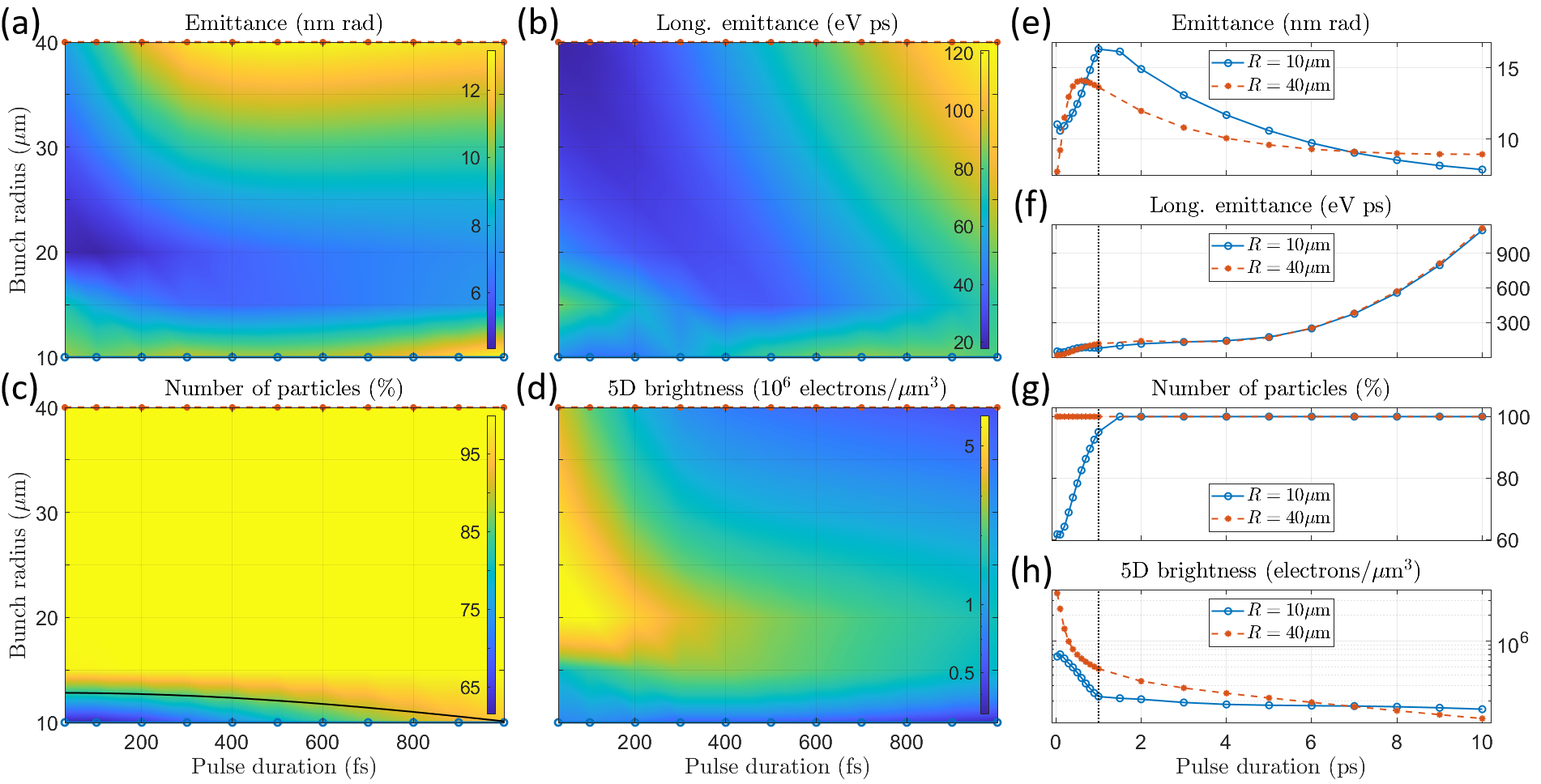}
\caption{Key output parameters of the RF photocathode gun depending on chosen values of the pulse duration and beam radius of the driving laser pulse. Plots in this figure are given for 0.16~pC bunch charge. \textbf{(a)} Transverse emittance; \textbf{(b)} Longitudinal emittance; \textbf{(c)} Fraction of particles exiting the RF gun, solid black line is the estimated onset of virtual cathode formation given by Eq.~\ref{Eq:virtual_cathode}; \textbf{(d)} Five-dimensional brightness of the extracted bunch; \textbf{(e-h)} Line cuts of colour plots (a-d) corresponding to the extreme values of the bunch radius, 10~$\mu$m and 40~$\mu$m. Note that these four graphs extend to larger values of the laser pulse duration, up to 10~ps. Dotted vertical line indicates the end value on the  colour plots (a-d).}
\label{fig:Astra_colormap}
\end{figure*}

Let us take a closer look at the simulation results in Fig.~\ref{fig:Astra_colormap}. We scan the initial radius and duration of the electron bunch and show four colour plots for its: (a) transverse emittance, (b) longitudinal emittance, (c) extracted bunch charge as a percentage of the intended value, and (d) 5D brightness. For applications such as UED and FEL, the 5D brightness is an important measure of beam quality. The gradual reduction of the extracted charge in Fig.~\ref{fig:Astra_colormap}c with decreasing bunch radii indicates the formation of a virtual cathode. Most importantly, in Fig.~\ref{fig:Astra_colormap}d we can clearly see the optimal aspect ratio of the bunch to reach the highest possible brightness. Note that the brightness is a factor of 20 higher than that for a long beam with a thermal emittance (Fig.~\ref{fig:Astra_colormap}h).

The panels e-h in Fig.~\ref{fig:Astra_colormap} show the 2D graphs of the bunch parameters for the  $10~\mu$m and $40~\mu$m radii, respectively. The limits of the abscissa coordinate are extended to much longer laser pulse duration to illustrate the trend towards thermal emittance. We note that before reaching the thermal asymptotic for very long bunches, the transverse emittance has a region of higher values around 1~ps, and the colour plots in Fig.~\ref{fig:Astra_colormap} cover the range up to that very time scale.
There is a clear transition between the two processes: (i) long bunch extraction with an almost thermal emittance (right-hand side of the graphs), and (ii) short ellipsoidal bunch blowout (left-hand side of the graphs). The maximum of the transverse emittance occurs because the extracted bunch is too long to become a uniformly filled ellipsoid with linear space-charge forces. The 1~ps ($300~\mu$m length) bunch duration is comparable or larger than the bunch radius ($10~\mu$m and $40~\mu$m in Figs.~\ref{fig:Astra_colormap} e-h). The electron bunch now has a shape of a cigar rather than a pancake. At the same time, the bunch is still too short to reach the thermal emittance of longer bunches. Thus, the high current density and nonlinear space-charge forces lead to rapid emittance degradation. This intermediate regime of extraction between (i) and (ii) should be avoided in order to achieve a high beam brightness.

\begin{figure*}[tb]
\centering
\includegraphics[width = \linewidth]{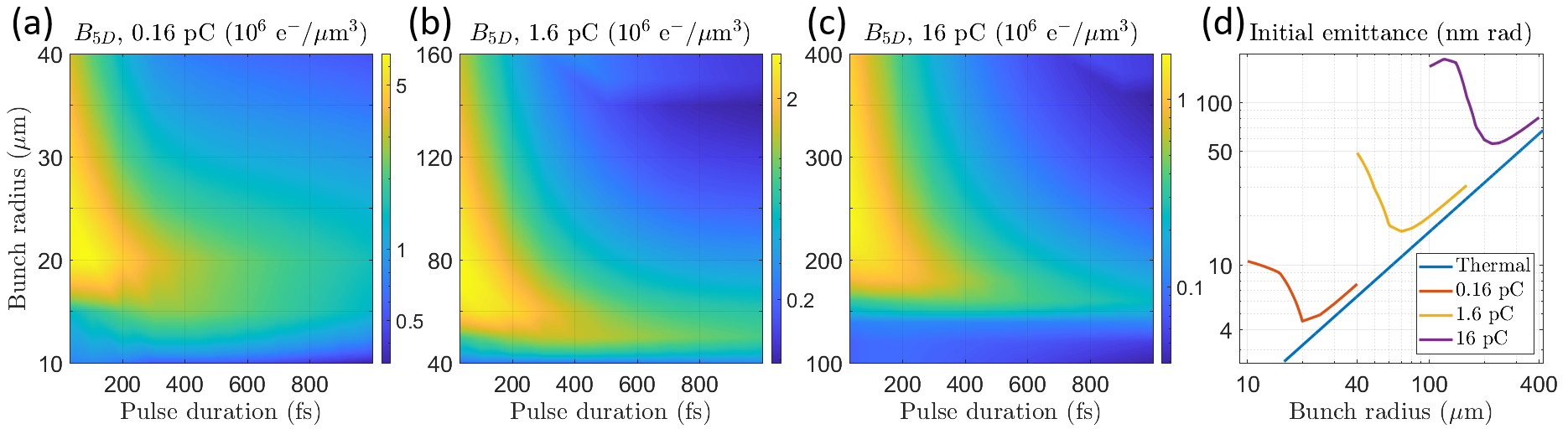}
\caption{\textbf{(a-c)} Five-dimensional brightness colour plots for the chosen values of the pulse duration and beam size of the driving laser pulse: \textbf{(a)} bunch charge of 0.16~pC, \textbf{(b)} bunch charge of 1.6~pC, \textbf{(c)} bunch charge of 16~pC. \textbf{(d)} Initial emittance of the extracted bunches depending on the bunch radius compared to the thermal emittance of a bunch of the same size; driving laser pulse duration is 30~fs.}
\label{fig:brightness}
\end{figure*}

In Fig.~\ref{fig:brightness}, a-c we look at how the bunch charge affects the 5D brightness of the extracted bunch. Clearly, with accounting for the bunch radius, the behaviour of the optimal point for brightness remains the same: a pancake-like bunch of the optimal radius provides the minimum emittance and thus maximum brightness. In addition, Fig.~\ref{fig:brightness}d shows that the emittance of 30~fs-long bunches approaches the thermal level, which is the minimum that can be attained physically. The output parameters in Figs.~\ref{fig:Astra_colormap}~and~\ref{fig:brightness} are read at a distance of 20 cm from the cathode, where the extracted bunch is to be injected into an emittance compensation coil or into a booster.

\begin{figure*}[tb]
\centering
\includegraphics[width = \linewidth]{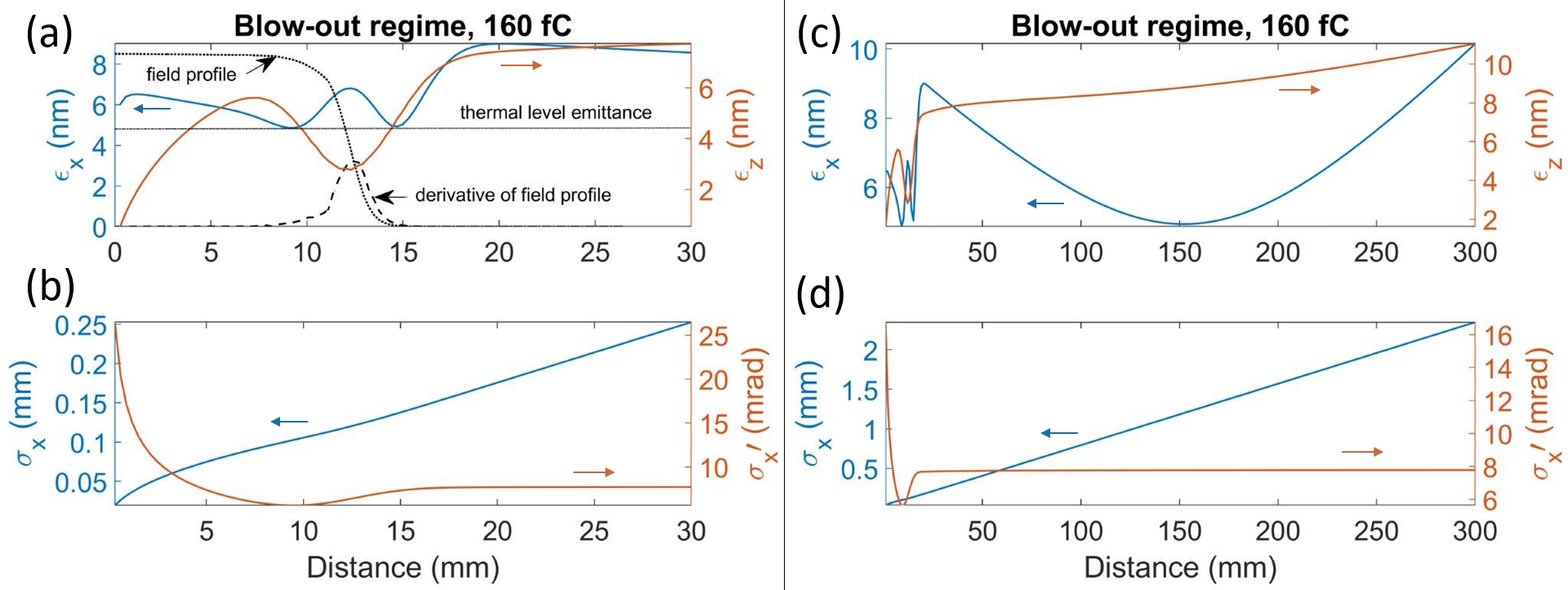}
\caption{Evolution of emittance [\textbf{(a)} and \textbf{(c)}] and beam moments [\textbf{(b)} and \textbf{(d)}] of the bunch generated in the blowout regime: \textbf{(a-b)} inside the RF gun and \textbf{(c-d)} after leaving it. For reader's convenience, thermal emittance, RF gun's field profile and its derivative are included in \textbf{(a)}. In the simulation, bunch charge of 0.16~pC, bunch radius of 25~$\mu$m and pulse duration of 30~fs were used (optimum in Fig.~\ref{fig:brightness}a).}
\label{fig:emittance_evolution}
\end{figure*}

\begin{figure*}[tb]
\centering
\includegraphics[width = \linewidth]{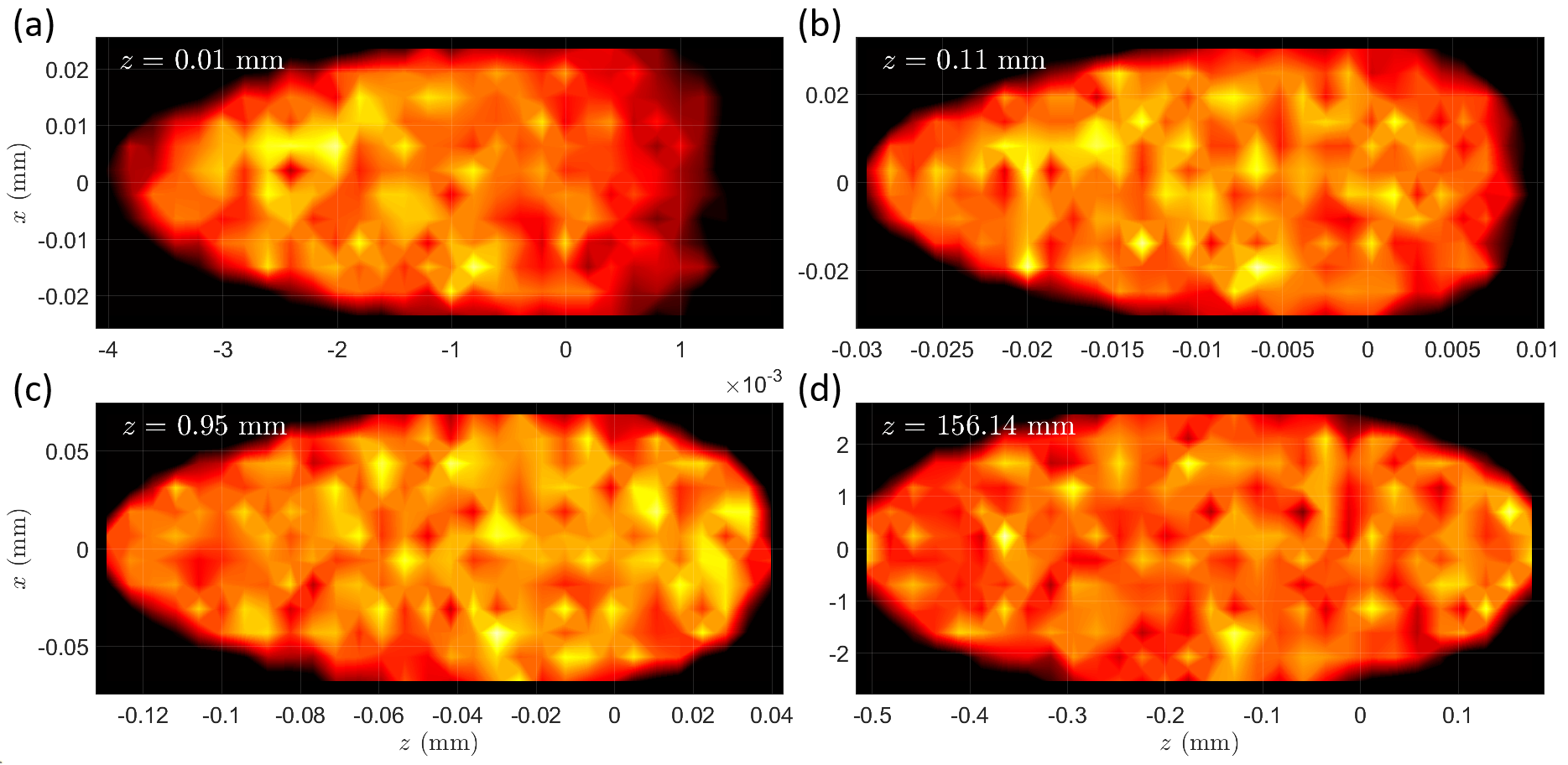}
\caption{Qualitative evolution of the charge density within the bunch as it propagates along the $z$-axis. Histograms are created in $x$-$z$ plane for a central slice of the bunch in the $y$-direction with a thickness of $0.15~\sigma_y$, which contains about 10\% of particles. In the simulation, bunch charge of 0.16~pC, bunch radius of 25~$\mu$m and pulse duration of 30~fs were used (optimum in Fig.~\ref{fig:brightness}a).}
\label{fig:density_evolution}
\end{figure*}

Let us choose the 5D brightness optimum in Fig.~\ref{fig:brightness}a, and consider how the bunch at these parameters evolves on its way out of the RF gun. The properties of interest are the transverse emittance $\epsilon_x$, longitudinal emittance $\epsilon_z$, transverse beam size $\sigma_x$ and divergence $\sigma_x'$, together determining the full 6D brightness of the bunch, see Fig.~\ref{fig:emittance_evolution}. As we see in Fig.~\ref{fig:emittance_evolution}, b and d, after exiting the gun through the aperture ($z \approx 15$~mm), the divergence becomes constant and the bunch expands linearly in the transverse plane.
The longitudinal emittance exhibits a minimum near the exit aperture corresponding to the maximum of the derivative of the accelerating field profile, Fig.~\ref{fig:emittance_evolution}a.

However, the most peculiar is the behaviour of the transverse emittance $\epsilon_x$. It has three minima at different $z$ values: before the exit aperture, right after it and out of the gun at $z \approx 15$~cm. 
At each point, the transverse emittance value approaches the thermal level. We stress that there is no emittance compensation line included in this simulation. We hypothesise that the complex emittance behaviour is the result of longitudinal plasma oscillations in the bunch during its blowout depicted in a qualitative way in Fig.~\ref{fig:density_evolution}.
Immediately after the emission, Fig.~\ref{fig:density_evolution}a ($z=10~\mu$m) the highest charge density is in the tail of the bunch. Closer to the exit aperture of the gun ($z\approx 1$~mm, Fig.~\ref{fig:density_evolution}c), the highest charge density is rather in the head of the bunch. Finally, upon approaching $z\approx 15$~cm (Fig.~\ref{fig:density_evolution}d) the charge density becomes uniform and the bunch ellipse non-warped.
Since the energy gain in the gun is on the level of 0.4~MeV, space-charge effects remain prominent, and the transverse and longitudinal planes are significantly coupled, leading to oscillatory exchange between $\epsilon_x$ and $\epsilon_z$, following the head-tail density oscillation. The latter might be initiated by the image charge on the cathode.
We also note that blowout generation of uniformly filled ellipsoidal bunches requires propagation over certain distance (about 15~cm in this case). A detailed study of emittance-related effects is under way.

\section{Onset of the virtual cathode}

In this section, we estimate the condition for the onset of a virtual cathode. Note that we use CGS units in this section.
As the electron bunch is extracted from the cathode during photoemission, a space-charge field builds up. It is composed of the fields of the bunch itself and its image on the cathode. The emission process diminishes to zero if the space-charge field reaches the level of the accelerating field – the onset of the virtual cathode. Let us estimate the space-charge field. We first consider a thin slice of the bunch with a slice charge $Q_{sl}$ and a characteristic radius R. Let this slice interact with another slice at a longitudinal distance $\Delta z$ away and the interaction energy be $W(\Delta z)$. Then, the longitudinal interaction force between the two slices is a gradient of the interaction energy $W$, and it turns out that the field acting on the slice (we call it an energy-wise field) can be written in the form~\cite{Shamuilov2018}
\begin{equation}
    E_{sl} = \frac{Q_{sl}}{4\pi R^2} \frac{\beta}{\left( 1 + \alpha |\Delta z|/R \right)^2},
\end{equation}
where the parameters $\alpha$ and $\beta$ depend on the form of the transverse density distribution. See~\cite{Shamuilov2018} for different distributions. For instance, for a homogeneous square-box distribution (cylindrical bunch) $\alpha \approx 1$ and $\beta \approx 2$. Then, for small distances from the slice, the energy wise field of the cylindrical bunch reads $E_{sl}=Q_{sl}/(2\pi R^2)$ and turns out to be equal to the on-axis field of a uniformly charged disk with the total charge $Q_{sl}$ – a well-anticipated result. 

To find the total space-charge field we need to convolute the energy-wise field of the slice with a longitudinal distribution of the slices $f(z)$. To find the maximum extractable charge, this distribution must be taken at the onset of the virtual cathode formation – the time of bunch emission corresponding to the balance between the space-charge field and the accelerating field so that the total field on the cathode drops to zero. Finding $f(z)$ analytically is a difficult task so we will rely on the observations from numerical simulations. First, for short bunches the onset of virtual cathode formation depends primarily on the already extracted total charge $Q_b$ and is almost independent of a specific temporal laser pulse shape – the effect already noticed by Luiten and co-workers~\cite{Luiten2004,vanOudheusden2010}. Second, for charges close to the maximum extractable charge, the longitudinal bunch density is quite uniform because of strong repulsion space-charge forces in the bunch. In other words, in the extracted bunch the slices are distributed approximately uniformly in the longitudinal direction with the transverse density profile still replicating the transverse laser pulse profile. This assumption contradicts findings shown in Fig.~\ref{fig:density_evolution}, but nonetheless allows to make a useful estimate.
Third, with a good accuracy the effective bunch length $L$ is $(e/2m)E_{acc}\Delta t^2$, where $\Delta t$  is the FWHM emission time. Then, for short bunches, $L<R$, the onset of virtual cathode formation occurs for
\begin{equation}\label{Eq:virtual_cathode}
    E_{acc} = \frac{\beta Q_b}{2\pi R (R + \alpha L)}.
\end{equation}
For a given accelerating field and a bunch charge, this equality gives the minimum bunch radius on the cathode as a function of the laser pulse duration.
In Fig.~\ref{fig:Astra_colormap}c, the curve given by this equation is the solid black line. Our extensive simulations indicate that the radius at which 5D brightness attains maximum for a given pulse duration is around 1.5 times of the critical radius. Thus, one can easily estimate the bunch parameters for maximising beam brightness.

\begin{table*}[tb]
\centering
\begin{tabular}{lcccccc}
     \hline \hline
     Parameter & Cornell & LCLS-II & CompactLight & Fig.~\ref{fig:brightness}c & Fig.~\ref{fig:brightness}b & Pegasus \\
     \hline
     Bunch charge (pC) & 19 & 20 & 16  &  16 & 1.6 & 1 \\
     Horizontal emittance (nm$\cdot$rad) & 230 & 250  & 54* & 55 & 15 & 40 \\
     Peak accelerating field (MV/m) & 8 & 34 & 120 & 35 & 35 & 70 \\
     \hline \hline
\end{tabular}
\caption{Comparison with state-of-the-art electron guns for key parameters. *calculated from the scaling law. }
\label{Table:comparison}
\end{table*}
\section{Comparison with other state-of-the-art guns}

An ideal candidate to compare our gun with is the heart of the CBETA energy-recovery linac photoinjector at the Cornell University (US). It is a DC electron gun that produces short electron bunches with energy about 400~keV and ultralow emittance~\cite{Gulliford2013,Gulliford2015}. 
Although of a different type, the bunch charge and energy it provides are very close to the output of our RF gun (Table~\ref{Table:comparison}).
The same range of these parameters is also listed in the LCLS-II injector specifications for the new APEX2 gun, the closest CW VHF relative. One can clearly see that for the same bunch charge, our RF gun can provide emittance values lower by a factor of 2 to 3. Another important parameter is the RF power consumption. For APEX2, the design value is 127~kW per cell, which is a factor of 6 larger than in our design. The earlier single-cell APEX gun consumes 85~kW, or four times more than our RF gun~\cite{Li2019}.

It is interesting to compare with the measurements of a cigar-like beam from the Pegasus laboratory at UCLA (US)~\cite{Li2012}. This process can be seen as transverse blowout, a string evolving into an ellipsoid, contrary to the longitudinal blowout of a pancake considered here. With a lower accelerating field and higher bunch charge, our RF gun can deliver a significantly lower emittance. However, R.~K.~Li~\textit{et al.} mention that measurements were limited, and a lower emittance can be achieved, according to simulations.

Finally, it is also interesting to compare our gun design with that of a 1-kHz, C-band photo-injector proposed for the  CompactLight project~\cite{CompactLight}. This project, financed by the European Union, is a consortium of 26 institutions around the world and aims at designing a compact and cost-effective X-band-driven FEL at soft and hard X-ray photon energies~\cite{CompactlightIPAC}. The nominal bunch charge in the CompactLight design is 75 pC but a scaling law~\cite{Simone_rf_gun} can be applied to estimate the emittance for the 16~pC case. It turns out that for an optimised photo-injector, the emittance for a sought bunch charge $Q_b$ scales as $\epsilon_\mathrm{ref} (Q_b/Q_\mathrm{ref})^{2/3}$, where $\epsilon_\mathrm{ref}$ and $Q_\mathrm{ref}$ are the reference (known) emittance and charge, respectively. 

Table~\ref{Table:comparison} clearly shows that our design of RF gun cavity competes with the state-of-the-art guns, including the superconducting alternatives~\cite{qian2019overview}. Naturally, ultralow values of emittance need to be preserved by using an emittance compensation line~\cite{Carlsten1989,Qiu1996}. Our preliminary simulations show that the standard emittance compensation works well and emittances below 50~nm are reachable. \newline



\section{Conclusion}

To sum up, we have presented a new APEX-like CW RF photo-gun with a high accelerating gradient and reduced power dissipation. It can operate in three different regimes: 160~fC, 100~fs bunches with a 5-nm-scale emittance for UED experiments, 1.6~pC, few-ps bunches with a 20-nm-scale emittance for table-top FELs and dielectric-based accelerators, and 16~pC, 10~ps bunches with a 50-nm-scale emittance for inverse Compton sources and other accelerator-based photon sources.

We have carried out detailed analysis of gun output for various bunch aspect ratios and found an optimal working point for blowout generation of high-brightness bunches to be at about 1.5 times of the critical radius $R$ found with Eq.~(\ref{Eq:virtual_cathode}). At the optimum, the generated bunches have emittance close to the thermal level and a brightness much higher than that of longer beams. We have compared the performance of our RF gun to that of state-of-the-art photoinjectors and found the design to be promising for user facilities.

The evolution of the blown-out bunch along the propagation axis, its emittance behaviour and charge density variations are examined. These subtle effects are a subject of a separate study and depend on a specific accelerator layout.

\bibliography{refs}
\bibliographystyle{unsrt}

\end{document}